\documentclass{ifacconf}

\usepackage{graphicx}      
\usepackage{natbib}        
\usepackage{algorithmic}
\usepackage{graphicx}
\usepackage{textcomp}
\usepackage{xcolor}
\usepackage{amssymb}
\usepackage[centertags]{amsmath}
\usepackage{relsize}
\usepackage{textfit}
\usepackage{dcolumn}
\usepackage{url}
\usepackage{tikz,pgfplots}
\usetikzlibrary{arrows,decorations.pathreplacing,intersections,calc}
\usepackage{comment} 
\usepackage{siunitx}
 \usepackage{url}

    \usepackage{graphicx}      
\usepackage{balance}        

\usepackage{algorithmic}
\usepackage{graphicx}
\usepackage{textcomp}
\usepackage{xcolor}
\usepackage{amssymb}
\usepackage[centertags]{amsmath}
\usepackage{relsize}
\usepackage{textfit}
\usepackage{dcolumn}
\usepackage{url}
\usepackage{tikz,pgfplots}
\usetikzlibrary{arrows,decorations.pathreplacing,intersections,calc}
\usepackage{comment} 
\usepackage{siunitx}
 \usepackage{url}
\usepackage{multirow}

\usepackage{tabularx}

\usepackage{exscale}

\usepackage{subfigure}
\usepackage{arydshln}

\usepackage{upgreek}

\usepackage{bm}

\usepackage{times,mathptmx}
\usepackage[T1]{fontenc}

\usepackage{cancel}

\newcommand\inlineeqno{\stepcounter{equation}\ (\theequation)}

\usepackage{exscale}

\usepackage{subfigure}
\usepackage{arydshln}

\usepackage{upgreek}

\usepackage{bm}

\usepackage{times,mathptmx}
\usepackage[T1]{fontenc}

\usepackage{cancel}

\begin{document}
\begin{frontmatter}

\title{Direct Integration of Recursive Gaussian Process Regression  Into Extended Kalman Filters With Application to Vapor Compression Cycle Control} 

\author[First]{Ricus Husmann} 
\author[First]{Sven Weishaupt} 
\author[First]{Harald Aschemann}

\address[First]{Chair of Mechatronics, University of Rostock, 
    Germany\\ \{Ricus.Husmann, Sven.Weishaupt, Harald.Aschemann\}@uni-rostock.de.}


\begin{abstract}                
This paper presents a real-time capable algorithm for the learning of Gaussian Processes (GP) for submodels. It extends an existing recursive Gaussian Process (RGP) algorithm which requires a measurable output. In many applications, however, an envisaged GP output is not directly measurable. Therefore, we present the integration of an RGP into an Extended Kalman Filter (EKF) for the combined state estimation and GP learning.  The algorithm is successfully tested in simulation studies and outperforms two alternative implementations -- especially if high measurement noise is present. We conclude the paper with an experimental validation within the control structure of a Vapor Compression Cycle typically used in refrigeration and heat pumps. In this application, the algorithm is used to learn a GP model for the heat-transfer values in dependency of several process parameters. The GP model significantly improves the tracking performance of a previously published model-based controller.
\end{abstract}


\end{frontmatter}

\section{Introduction}
Many methods are available to identify parametric models or uncertain model parts in an online manner. One common method is to use parametric functions that are linear in certain parameters, e.g. polynomial ansatz functions, to apply  recursive least-squares techniques as described in \cite{Blum:1957}. This may also be extended towards linear Kalman Filters (KF) or Unscented/Extended Kalman Filter (UKF/EKF) implementations if certain system states are not measurable, see \cite{Kalman:1960}, \cite{Julier:1997}. Another option is given by Moving Horizon Estimation (MHE), which can address state as well as input constraints and may be advantageous for nonlinear systems, cf.~\cite{Haseltine:2005}. A more general approach involves an online training of feedforward neural networks as proposed in \cite{Jain:2014}. While enabling very accurate fits, neural networks contain many weights and are, hence, prone to overfitting.

As they offer flexibility, robustness against overfitting and a built-in uncertainty measure, Gaussian Processes (GP) represent a powerful tool for data-driven modeling, see \cite{Schuerch:2020}. An example for the achievable performance of GP-based methods in comparison to classical observer-based disturbance compensation is given by \cite{Aschemann:2020}. Classic GP regression corresponds to a non-parametric model and suffers from significant limitations when it comes to an online implementation. First, additional data increases the model complexity, which is undesirable for real-time purposes. Second, the computational cost of evaluating a model includes a matrix inversion and grows with $O(n^3)$, which limits the data size to a few thousand data points only. Many methods are described in the literature that address this issue. Sparse approximations like the active set-type methods constrain the use of training data points to reduce the computational burden as described in \cite{Quinonero:2005}. A promising approach has been published in \cite{Huber:2013} with a recursive Gaussian Process regression (RGP). The idea is to select grid points based on basis vectors, where the GP kernels are updated in real time with new data. With the use of basis vectors, the Gaussian Process can now be interpreted as a parametric model. This maintains most of the benefits of GP regression but leads to relatively small computational costs for the online evaluation.

Similar to traditional GP regression, the RGP algorithm utilizes a measured output with an uncertainty description that the GP model has to learn to make future predictions. Often, however, the output is not directly measurable. Since the RGP has a structure similar to a KF, it seems reasonable to use a KF to predict this output. In \cite{RGP_KF:2024}, we utilize an auxiliary parameter as a link between a KF and an RGP. In this paper, however, we present a seamless integration of the RGP into an EKF.

A motivating example for the development of the algorithm is given by the need for an identification of the heat-transfer value of a Vapor Compression Cycle (VCC) evaporator. In previous work of the authors, an integral feedback control by means of the heat-transfer value proved to be advantageous within a partial Input-Output Linearisation (IOL), cf.~\cite{VCC_Part:2024}. There, the possible benefit of learning a model for the heat-transfer value was pointed out.

The paper is organized as follows: after an introduction of the original RGP algorithm and some adaptations regarding the implementation in Sect.~\ref{sec:RGP_integration}, we introduce the problem and the KF integration in Sect.~\ref{sec:KF_integration} that results in the RGB-dKF algorithm. Here, different steps of the integration are presented, including an RGP integration for pure feedforward evaluation. In Sect.~\ref{sec:sim_val}, we present a comparative analysis in simulation with a baseline solution and the results presented in \cite{RGP_KF:2024}. Then, an implementation of the RGP-dKF in combination with a partial IOL controller at a test rig of a VCC is described together with a validation by means of experimental results. The paper closes with conclusions and an outlook.

\section{Recursive Gaussian Process Regression}
\label{sec:RGP_integration}

If the following, we will give a short introduction into the recursive Gaussian Process regression (RGP) algorithm as in \cite{Huber:2013} and present some adaptations. As in most GP applications, we employ a Squared Exponential (SE) kernel
\begin{equation}
k(\bm{x},\bm{x}')=\sigma_K^2 \cdot \exp(-(\bm{x}-\bm{x}')^T(\bm{x}-\bm{x}')(2 L)^{-1}) \,
\end{equation}
with a zero mean function $m(\bm{x})=0$. The hyperparameters $L$ and $\sigma_K$ are defined by the user. As explained in Subsec. \ref{sec:Inp_Norm}, we define a single $L$ for all inputs and consider the different input ranges by normalization. Since we envisage a real-time implementation, only one measurement per time step $k$ is available, which leads to $n_y=1$. In the following, $\bm{X}_k$ are the test inputs at time step $k$ and $\bm{X}$ denotes the collection of so called basis vectors, which are defined by the user, see Subsec. \ref{sec:Inp_Norm}. The variable $y_{GP,k}$ is the measurement of the output that is to be predicted by the GP, and $\sigma_{y,GP}^2$ denotes its corresponding measurement variance. $\bm{\mu}^g_{z,k}$ and $\bm{C}_{z,k}^g$ represent the mean values and the covariance of the kernel, respectively.

The precomputed variables and the initialization of the kernel states
\begin{center}
\begin{tabularx}{1.028\columnwidth}{l X r} 
  \multirow{4}{1em}{\rotatebox{90}{{Offline}}} & $\bm{K}=k(\bm{X},\bm{X})$~,  &$\inlineeqno$\\ 
& $\bm{K}_{I}=\bm{K}^{-1}$~,& \\ 
& $\bm{\mu}_{z,0}^g= \bm{0}$~,&\\ 
& $\bm{C}_{z,0}^g= \bm{K}$~,&\\ 
\end{tabularx}
\end{center}
are defined as in  \cite{Huber:2013}. With a zero mean and a scalar measurement, the inference step from \cite{Huber:2013} simplifies to 
\begin{center}
\begin{tabularx}{1.028\columnwidth}{l X r} 
  \multirow{3}{1em}{\rotatebox{90}{{Inference}}} & $\bm{J}_{k} =k(\bm{X}_k,\bm{X}) \cdot \bm{K}_{I}$~,&$\inlineeqno$\\ 
& $\bm{\mu}_{z,k}^p=\bm{J}_k \cdot \bm{\mu}^g_{z,k}$~,& \\ 
& $\bm{C}_{z,k}^p=\cancelto{\sigma_K^2}{k(\bm{X}_k,\bm{X}_k)}+\bm{J}_k (\bm{C}_{z,k}^g-\bm{K}) \bm{J}_k^T$~.\\  \label{eq:RGP_Inference}
\end{tabularx}
\end{center}
Here, the superscript $p$ denotes the prediction of the GP  for the test inputs $\bm{X}_k$. As a first modification, we introduce process noise $\sigma_p$ into the update step
\begin{center}
\begin{tabularx}{1.028\columnwidth}{l X r} 
  \multirow{3}{1em}{\rotatebox{90}{{Update}}} & $\tilde{\bm{G}}_{z,k}=\bm{C}_{z,k}^g \bm{J}_k^T \cdot(\bm{C}_{z,k}^p+ \sigma_{y,GP}^2 )^{-1}$~, &$\inlineeqno$\\ 
& $\bm{\mu}_{z,k+1}^g= \bm{\mu}_{z,k}^g+\tilde{\bm{G}}_{z,k} \cdot (y_{GP,k}- \bm{\mu}_{z,k}^p)$~,& \\ 
& $\bm{C}_{z,k+1}^g=\bm{C}_{z,k}^g-\tilde{\bm{G}}_{z,k} \bm{J}_k\bm{C}_{z,k}^g+\bm{I} \sigma_p^2$~. \\ 
\end{tabularx}
\end{center}
This process noise can be considered similar to a forgetting factor in recursive least-squares parameter estimation and allows for the consideration of changes in the hidden function. If $\sigma_p^2>0$, a bounding of $\bm{C}_{z,k}^g$ may be useful.

In the following subsections, some further adaptations regarding the implementation are presented.

\subsection{Input Normalization} 
\label{sec:Inp_Norm}
In this paper, the basis vectors are defined as an equidistant grid with distance $1$ along all input axes. Thus, the so called basis vectors are collected in a $(\prod_{i=1}^{n_{X}} N_i) \, \times \, n_{X}$-dimensional matrix that includes all  vertices in the grid. Here, $n_{X}$ denotes the input dimension of the GP and $N_i$ the number of basis vector points (BVP) in the respective dimension $i$. For $n_X=2$ , $N_1=2$ and $N_2=3$  exemplary basis vectors would be 
\begin{equation}
\bm{X}=\left[\begin{matrix}  \bm{X}_1^T\\ \bm{X}_2^T\end{matrix}\right]^T=\left[\begin{matrix}  0 &1 &0&1&0&1 \\  0 &0 &1&1&2&2 \end{matrix}\right]^T \,.
\end{equation}
The biggest advantage of the standardized grid definition is that the choice of $L$  is simplified significantly.  Now, for many problems the same basis vector- and $L$-definition can be applied, and a single $L$ for all dimensions becomes a reasonable choice. 
Naturally, the inputs of the GP $\zeta_{i}$ are related to physical values. Hence, a normalization 
\begin{equation}
X_{i,k}=f_{norm}(\zeta_{k,i})=(\zeta_{i,k}-\underline{\zeta}_{i}) \cdot \frac{N_i-1}{\overline{\zeta}_{i}-\underline{\zeta}_{i}}  \,,
\end{equation}
is applied for each input before each GP evaluation, where $\underline{\zeta}_{i}$ and $\overline{\zeta}_{i}$ denote the lower and upper bound of the input respectively. 

\subsection{Improved Numerical Stability}
\label{sec:numstab}
%
%
%
In the original algorithm, the inversion of the kernel matrix ${\bm{K}_{I}=\bm{K}^{-1}}$ is computed offline for a later use in the online computation of $\bm{J}_k=k(\bm{X}_k,\bm{X}) \bm{K}_{I}$. Naturally, this has great benefits in terms of computational speed. It was, however, discovered that even moderate values of $L$ may lead to an almost singular $\bm{K}$ and, thus, an ill-conditioned inversion, which greatly deteriorates the performance of the RGP. While the straightforward online solution of the linear equation $\bm{J}_k \bm{K}=k(\bm{X}_k,\bm{X})$ solves the numerical issues, it is computationally much more expensive. A remedy is provided by an offline QR decomposition ${[\bm{Q},\bm{R}]=qr(\bm{K})}$, where $\bm{Q}$ is orthogonal and $\bm{R}$ is upper triangular, followed by a subsequent online solution of the linear equation $ \bm{J}_k \bm{R}^T = k(\bm{X}_k,\bm{X})\bm{Q}$ leveraging the triangular structure of $R$. This preserves the numerical advantages, while the relative computational cost can be reduced to a factor $T_{QR}/T_{I}\approx 2$ -- independent of the basis vector dimension. The proposed method is abbreviated by $\bm{J}_k=k(\bm{X}_k,\bm{X})/ \bm{K}$ in the following.

\section{Kalman Filter Integration}
\label{sec:KF_integration}

\subsection{Problem Formulation}
We consider nonlinear discrete-time systems 
\begin{equation}
\bm{x}_{k+1}=\bm{f}(\bm{x}_k,\bm{u}_k,z_k(\bm{\zeta}_k)) \,, \bm{y}_k=\bm{h}(\bm{x}_k)  	\label{eq:LinDiffEq}
\end{equation}
with states $\bm{x}_k$ and inputs $\bm{u}_k$, the measurement output $\bm{y}_k$ and the disturbance $z_k$, which depends in a nonlinear manner on a deterministic signal $\bm{\zeta}_k$. 


\subsection{Pure GP Prediction}
\label{sec:KF_GP_pred}
In the case that the GP model is purely used for prediction of a disturbance within an EKF setting, only the inference part of the RGP algorithm has to be evaluated
\begin{align}
\bm{X}_k&=f_{norm}(\bm{\zeta}_k) \,, \\ \nonumber
\bm{J}_{k} &=k(\bm{X}_k,\bm{X})/ \bm{K} \,, \\ \nonumber
\bm{\mu}_{z,k}^p&=\bm{J}_k \cdot \bm{\mu}^g_{z,k} \,, \\ \nonumber
\bm{C}_{z,k}^p&=\sigma_K^2+\bm{J}_k (\bm{C}_{z,k}^g-\bm{K}) \bm{J}_k^T~ . \label{KF_RGP_Inference}
\end{align}
A GP model with finite basis vector length is not able to fully reconstruct any non-trivial hidden function. Moreover, numerical errors need to be considered. To account for this uncertainty, we introduce an additional $\sigma_r$ into the GP prediction $\bm{\tilde{C}}_{z,k}^p=\bm{C}_{z,k}^p+\sigma_r^2 $. This value $\sigma_r$ will generally be relatively small and provides a lower bound for the GP prediction uncertainty. In high-noise environments or if $\sigma_p \gg 0$ holds, it has no significant impact. In the low-noise case it can, however, prevent from an underestimation of the GP noise.

The predicted disturbance $\bm{\tilde{C}}_{z,k}^p$ can now be integrated into the EKF as a noisy input
\begin{align}
\bm{\mu}_{x,k+1}^p&=\bm{f}(\bm{\mu}_{x,k}^g,{\mu}_{z,k}^p,u_k) \,, \\ \nonumber
\bm{C}_{x,k+1}^p&={\bm{A}}_k \bm{C}_{x,k}^g  {\bm{A}}_k^T +\bm{Q}_x +  {\bm{e}}_k\bm{\tilde{C}}_{z,k}^p {\bm{e}}_k^T~,
\end{align}
with the linearized system matrix and the disturbance input vector
\begin{equation}
{\bm{A}}_k=\left.\frac{\partial \bm{f}}{\partial\bm{x}_k}\right|_{\bm{\mu}_{x,k}^g,{\mu}_{z,k}^p,u_k}  \,,~~ 
{\bm{e}}_k=\left.\frac{\partial \bm{f}}{\partial\bm{z}_k}\right|_{\bm{\mu}_{x,k}^g,{\mu}_{z,k}^p,u_k}  \,, 
\end{equation}
within the covariance prediction. The corresponding update is identical to the standard EKF update
\begin{align}
{\bm{H}}_k&=\left.\frac{\partial \bm{h}}{\partial\bm{x}_k}\right|_{\bm{\mu}_{x,k+1}^p} ~,\\  \nonumber
{\bm{G}}_{x,k}&=\bm{C}_{x,k+1}^p {\bm{H}}_k^T \cdot({\bm{H}}_k  \bm{C}_{x,k+1}^p{\bm{H}}_k^T+ \bm{R}_x)^{-1} ~, \\  \nonumber
 \bm{\mu}_{x,k+1}^g&= \bm{\mu}_{x,k}^g+{\bm{G}}_{x,k} \cdot (\bm{y}_k- \bm{h}(\bm{\mu}_{x,k+1}^p)) ~, \\ \nonumber
 \bm{C}_{x,k+1}^g&=\bm{C}_{x,k}^g-{\bm{G}}_{x,k} {\bm{H}}_k \bm{C}_{x,k+1}^p ~ .
\end{align}

\subsection{GP Prediction and Training}
Now, we want to introduce the full simultaneous update of system states and RGP states. In a first step, we add the GP kernel mean values $\bm{\mu}_{z,k}$ to the state vector, which yields the extended mean value vector $ \bm{\mu}_{k}=[\bm{\mu}_{x,k}^T,\bm{\mu}_{z,k}^T]^T$. Again, the nonlinear mean value prediction is not affected by this measure, besides the fact that $\bm{\mu}_{x,k}^g=\bm{\mu}_{k}^g(1:n_x)$ and $\bm{\mu}_{z,k}^g=\bm{\mu}_{k}^g(n_x+1:end)$ hold. The full covariance prediction becomes
\begin{equation}
\bm{C}_{k+1}^p=\bm{\tilde{A}}\bm{C}_{k}^g\bm{\tilde{A}}^T+\bm{\tilde{e}}_k \left(\sigma_K^2-\bm{J}_k \bm{K} \bm{J}_k^T +\sigma_r^2\right)\bm{\tilde{e}}_k^T+\bm{Q}~,
\end{equation}
with
\begin{equation}
\bm{\tilde{A}}_k=\left[\begin{matrix} {\bm{A}}_k &  {\bm{e}}_k \bm{J}_k\\ \bm{0} & \bm{I} \end{matrix}\right] \, ,  \, \bm{\tilde{e}}_k=\left[\begin{matrix} \bm{e}_k \\ \bm{0} \end{matrix}\right]\, \mathrm{and} \,\, \bm{Q}=\left[\begin{matrix} \bm{Q}_x &  \bm{0}\\ \bm{0} & \bm{I} \sigma_p^2 \end{matrix}\right] \label{eq:KF_GP_def} \,.
\end{equation}
The extension of the system matrix simply marks the linearized prediction equations of the RGP, whereas the additional input term $\sigma_K^2-\bm{J}_k \bm{K} \bm{J}_k^T+ \sigma_r^2$ completes the prediction noise of the GP. For the special case of zero covariances between system states and GP states, this integration simplifies to the pure GP prediction in Subsec. \ref{sec:KF_GP_pred}. Please note the direct integration of the GP process noise into the overall process noise in  \eqref{eq:KF_GP_def}.

For the following EKF update with simultaneous RGP training, we only need to introduce the extended measurement vector  
\begin{equation}
\bm{\tilde{H}}_k=\left[\bm{H}_k, \bm{0} \right] \,.
\end{equation}
The rest of the update corresponds to the standard EKF update in Subsec. \ref{sec:KF_GP_pred} -- however, with the extended mean value vector and covariance matrix.

\subsection{Summary of the RGP-dKF algorithm}
In the following, the complete algorithm with integrated RGP prediction and update (RGP-dKF) is summarized. It may be beneficial to reject an RGP update and to update the system states only. This is done in variant 4b, which summarizes Subsec. \ref{sec:KF_GP_pred}. 

\newpage
First, initialize 
\begin{center}
\begin{tabularx}{1.028\columnwidth}{l X r} 
  \multirow{6}{1em}{\rotatebox{90}{{Offline}}} & $\bm{K}=k(\bm{X},\bm{X})$~,  &$\inlineeqno$\\ 
& $\bm{\mu}_{0}^g=\left[\bm{x}_0^T,\bm{0}\right]^T$~, &\\ 
&&\\
& $\bm{C}_{0}^g=\left[\begin{matrix}\bm{C}_{x,0}^g &\bm{0} \\ \bm{0} &\bm{K}   \end{matrix}\right]$~. &\\ 
&&\\
\end{tabularx}
\end{center}
Now, evaluate the following recursive algorithm for all steps $k=0,1,2,..$: \newline
1. Complete the RGP inference with $\bm{\mu}_{x,k}^g=\bm{\mu}_{k}^g(1:n_x)$ and $\bm{\mu}_{z,k}^g=\bm{\mu}_{k}^g(n_x+1:end)$
\begin{center}
\begin{tabularx}{1.028\columnwidth}{l X r} 
  \multirow{4}{1em}{\rotatebox{90}{{Inference}}} & $\bm{X}_k=f_{norm}(\bm{\zeta}_k)$~, &$\inlineeqno$\\ 
& $\bm{J}_k=k(\bm{X}_k,\bm{X}) / \bm{K}$~,& \\ 
& $ \mu_{z,k}^p=\bm{J}_k \cdot \bm{\mu}^g_{z,k}$~. &\\ 
\end{tabularx}
\end{center}
2. Linearize and evaluate
\begin{center}
\begin{tabularx}{1.028\columnwidth}{l X r} 
  \multirow{4}{1em}{\rotatebox{90}{{Linearization}}} & $\bm{\tilde{A}}_k=\left[\begin{matrix}  \left.\frac{\partial \bm{f}}{\partial\bm{x}_k}\right|_{\bm{\mu}_{x,k}^g,{\mu}_{z,k}^p,u_k}&  \left.\frac{\partial \bm{f}}{\partial\bm{z}_k}\right|_{\bm{\mu}_{x,k}^g,{\mu}_{z,k}^p,u_k} \bm{J}_k\\ \bm{0} & \bm{I} \end{matrix}\right]$~, & $\inlineeqno$\\ 
& $\bm{\tilde{e}}_k=\left[\begin{matrix}  \left(\left.\frac{\partial \bm{f}}{\partial\bm{z}_k}\right|_{\bm{\mu}_{x,k}^g,{\mu}_{z,k}^p,u_k}\right)^T &\bm{0} \end{matrix}\right]^T$~. & \\ 
\end{tabularx}
\end{center}
3. Predict the extended system
\begin{center}
\label{eq:KF_eq}
\begin{tabularx}{1.028\columnwidth}{l X r} 
\multirow{3}{1em}{\rotatebox{90}{KF-Pred.}} & $\bm{\mu}_{k+1}^p=\bm{f}(\bm{\mu}_{x,k}^g,{\mu}_{z,k}^p,u_k) $~, &$\inlineeqno$\\ 
& $\bm{{C}}_{k+1}^p=\bm{\tilde{A}}\bm{{C}}_{k}^g\bm{\tilde{A}}^T+\bm{Q}+$ & \\ & \hspace{1cm}  $\bm{\tilde{e}}_k \left(\sigma_K^2-\bm{J}_k \bm{K} \bm{J}_k^T +\sigma_r^2\right)\bm{\tilde{e}}_k^T$~.& \\ 
\end{tabularx}
\end{center}
4a. State and RGP update with $\bm{\mu}_{x,k+1}^p=\bm{\mu}_{k+1}^p(1:n_x)$
\begin{center}
\begin{tabularx}{1.028\columnwidth}{l X r} 
  \multirow{5}{1em}{\rotatebox{90}{{Update}}}& ${\bm{\tilde{H}}}_k=\left[\left.\frac{\partial \bm{h}}{\partial \bm{x}_k}\right|_{\bm{\mu}_{x,k+1}^p},\bm{0}\right] 
$~,&$\inlineeqno $ \\
& $\tilde{\bm{G}}_{k}=\bm{C}_{k+1}^p \tilde{\bm{H}}_k^T \cdot(\tilde{\bm{H}}_k  \bm{C}_{k+1}^p\tilde{\bm{H}}_k^T+ \bm{R}_x)^{-1}$~, & \\ 
&$ \bm{\mu}_{k+1}^g= \bm{\mu}_{k}^g+\tilde{\bm{G}}_{k} \cdot (\bm{y}_k- \bm{h}( \bm{\mu}_{x,k+1}^p))$~, & \\
& $ \bm{C}_{k+1}^g=\bm{C}_{k}^g-\tilde{\bm{G}}_{k} \tilde{\bm{H}}_k \bm{C}_{k+1}^p$~.& \\
\end{tabularx}
\end{center}
4b. Pure state update with $\bm{\mu}_{x,k+1}^p=\bm{\mu}_{k+1}^p(1:n_x)$, $\bm{C}_{x,k+1}^p=\bm{C}_{k+1}^p(1:n_x,1:n_x)$,  $\bm{E}=\left[ \begin{matrix}\bm{I}\\\bm{0} \end{matrix}\right]$
\begin{center}
\begin{tabularx}{1.028\columnwidth}{l X r} 
  \multirow{5}{1em}{\rotatebox{90}{{Update}}}& ${\bm{H}}_k=\left.\frac{\partial \bm{h}}{\partial\bm{x}_k}\right|_{\bm{\mu}_{x,k+1}^p}
$~, &$\inlineeqno $ \\
& ${\bm{G}}_{k}=\bm{C}_{x,k+1}^p {\bm{H}}_k^T \cdot({\bm{H}}_k  \bm{C}_{x,k+1}^p{\bm{H}}_k^T+ \bm{R}_x)^{-1}$~, & \\ 
&$ \bm{\mu}_{k+1}^g= \bm{\mu}_{k+1}^p+\bm{E}{\bm{G}}_{k} \cdot (\bm{y}_k- \bm{h}( \bm{\mu}_{x,k+1}^p))$~, & \\
& $ \bm{C}_{k+1}^g=\bm{C}_{k+1}^p-\bm{E}{\bm{G}}_{k} {\bm{H}}_k \bm{C}_{x,k+1}^p \bm{E}^T$~. & \\
\end{tabularx}
\end{center}

\subsection{Discussion}
This algorithm can now be interpreted as an EKF implementation of the RGP with an exact propagation of the mean and covariance w.r.t.~the GP prediction. The directly measurable GP output -- as in the standard RGP formulation in Sec. \ref{sec:RGP_integration} -- can be reproduced by setting $\bm{A}_k=\bm{0}$ and $\bm{H}_k=({\bm{e}}_k\neq 0)^T$. In that particular case, the algorithm is an exact reformulation of the RGP algorithm, which can be shown in simulation studies up to numerical accuracy.

Since the hidden function and, hence, the remaining uncertainty of the GP model is usually unknown, $\sigma_r$ may be hard to obtain. Additionally, it depends on the particular choice for the GP hyperparameters. As a rule of thumb, $\sigma_r=0.01...0.1 \sigma_K$ proved to be a useful starting point. For linear time-invariant systems, the noise through $\sigma_r$ can be considered by means of an extension of the process noise.


Giving any guarantees of the performance of the RGP-dKF-algorithm is -- as with all nonlinear filter variants -- a very hard task. Possibly, conclusions may be drawn by means of observability Gramians as in \cite{Conley:2000} or \cite{Krener:2009}. This is, however, still the subject of on-going research.

As common practice with GPs, a simultaneous consideration of multiple disturbances $\bm{z}_k$ can be achieved by means of a repeated implementation of the RGPs.

\section{Simulative validation}
\label{sec:sim_val}
A second-order system
\begin{equation}
\dot{\bm{x}}=\left[\begin{matrix} 0 & 1 \\-4 &-4 \end{matrix} \right] \bm{x}+ \left[\begin{matrix} 0  \\1 \end{matrix} \right] u + \left[\begin{matrix} 0  \\1 \end{matrix} \right]   z
\end{equation}
with a disturbance $z(t)=-10 (1+0.1 \zeta(t)+\zeta(t)^3) $ is employed as an introductory example to validate the presented algorithm.  In the following, we investigate configurations with either one or two measurable states
\begin{equation}
{y}_1=\left[\begin{matrix} 1 &0\end{matrix} \right] \bm{x} \,,~ {\bm{y}}_2=\left[\begin{matrix} 1 &0\\0&1 \end{matrix} \right] \bm{x}    \,.
\end{equation}
Moreover, Gaussian white noise is added to these measurements. $u(t)$ is given by a sequence of step functions, whereas $\zeta(t)$ represents normally distributed colored noise. The system is discretized using the explicit Euler method. The algorithms are implemented according to Subsec.~\ref{sec:baseline} and \ref{sec:KF_integration}, respectively. All hyperparameters, except  $\sigma_{y,GP}$ for the baseline solution, are kept identical between all algorithms. The process noise of the RGP is set to zero, i.e. $\sigma_p=0$.

\subsection{Alternative Solutions}
\label{sec:baseline}
For linear systems -- if all states are measurable -- the disturbance can be calculated by inverting the finite-difference equations \eqref{eq:LinDiffEq} for the measurement $y_{GP,k}$ of the disturbance $z$, e.g. $y_{GP,k}=\bm{e}_k^{+} (\bm{x}_{k+1}-(\bm{A}_k \bm{x}_k+ \bm{B}_k \bm{u}_k))~, $ with the pseudo-inverse $\bm{e}_k^{+}=(\bm{e}_k^T \bm{e}_k)^{-1} \bm{e}_k^T$. We will use this approach as a baseline to calculate the measurement output for the RGP algorithm presented in Sec.~\ref{sec:RGP_integration} (RGP-B). A propagation of the uncertainties resulted in extremely small learning rates in this case. Therefore, $\sigma_{y,GP}$ is interpreted as a design parameter and is optimized separately for each case.

A second solution is given by an alternative implementation (RGP-KF), see (\cite{RGP_KF:2024}). This algorithm involves an integration of an RGP into a KF -- however, with an auxiliary uncertain parameter as a connection between the two algorithms. Since that work is still under review, a thorough comparison is not provided. Nevertheless, for a first comparison, the performance measures of that variant are presented here as well.

\subsection{Results}
\begin{figure}
	 \begin{center}
	 \vspace{-0.2cm}
		 \includegraphics[width=1\linewidth]{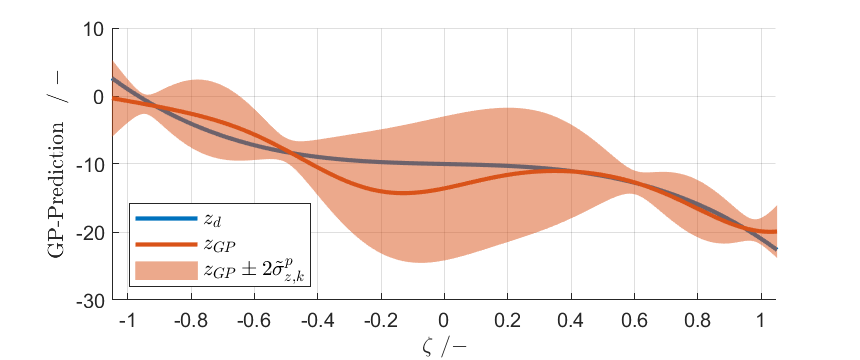}
		  \caption{GP prediction and $\pm 2\sigma$-range $0.5s$ after enabling the GP training.} 
		  \label{pic:Sim_plot05s}
	 \end{center}
\end{figure}

\begin{figure}
	 \begin{center}
	 	 \vspace{-0.2cm}
		 \includegraphics[width=1\linewidth]{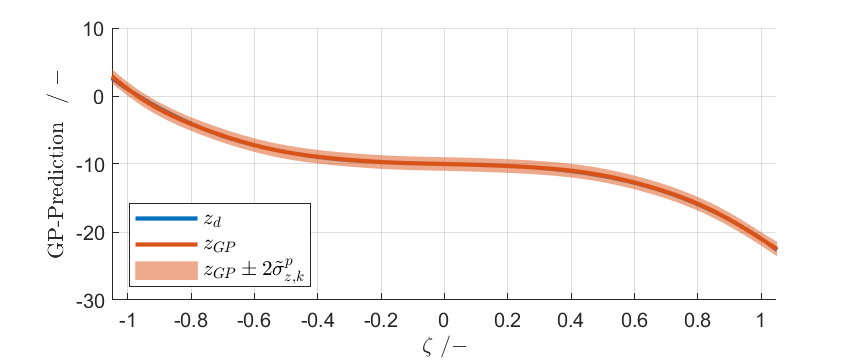}
		  \caption{GP prediction and $\pm 2\sigma$-range $100s$ after enabling the GP training.} 
		  \label{pic:Sim_plot100s}
	 \end{center}
\end{figure}

Figs.~\ref{pic:Sim_plot05s} and \ref{pic:Sim_plot100s} illustrate the resulting GP models and the hidden function for the RGP-dKF algorithm with $\bm{y}=\bm{y}_2$ and low measurement noise after $0.5~s$ and $100~s$ training time, respectively, together with their $2\sigma$-ranges. In Fig.~\ref{pic:Sim_plot05s}, it becomes obvious that regions of the hidden function, where more measurements are available, are characterized by a small uncertainty and a significantly better model fit. Overall, the hidden function stays within the $2\sigma$-range. In the second plot, Fig.~\ref{pic:Sim_plot100s}, the GP has almost converged and reflects the hidden function nearly perfectly -- which is to be expected for a linear system with low measurement noise. Here, the remaining covariance mainly depends on the choice of $\sigma_r$, which was chosen rather conservatively.

In Table \ref{table:perf}, two performance measures each are stated for the following three scenarios: high noise  $\bm{y}=\bm{y}_2$ (S1), low noise  $\bm{y}=\bm{y}_2$ (S2) and low noise  $\bm{y}=\bm{y}_1$ (S3). The first performance measure is the Root Mean Square Error (RMSE) of the GP prediction w.r.t. the ground truth,  which provides a measure for the mean value prediction. The second performance measure is the negative logarithmic likelihood (NLL), which is a popular way to quantify how well a filter can estimate both the mean value and the uncertainty, see \cite{huber:2015}. For both measures, a low value is desirable. It becomes obvious that the RGP-dKF algorithm shows the best performance in terms of RMSE for all three scenarios. In terms of the NLL, the RGP-dKF compares favorably in the S1 and S3. Only in the low-noise scenario S2, the alternatives performed better, but the performance of the RGP-dKF is still acceptable. These results might have been affected by a rather conservative choice of $\sigma_r$. 

 \begin{table}
\centering
  \caption{RMSE and NLL for the alternative RGP implementations and test scenarios.} 
\begin{tabular}{ l|c|c|c|c|c|c } 
   & \multicolumn{3}{c|}{RMSE}  & \multicolumn{3}{c}{NLL}     \\
    & S1  & S2 &  S3 & S1  & S2 &  S3   \\
\hline
$RGP-B$  & $1.30 $ &  $0.14$ &-&$ 2.5 $&  $\bm{-0.77}$&-\\
 \hline
$RGP-KF$& $ 0.71 $ & $0.12$&$0.84$& $1.87$  & $-0.41$&$2.49$\\
 \hline
 $RGP-dKF$& $ \bm{0.32}$  & $\bm{0.10}$&$\bm{0.46}$&  $\bm{1.19}$  & $1.16$&$\bm{1.18}$\\
 \end{tabular}
\label{table:perf}
\end{table}

\section{Experimental Results for a VCC}
\label{sec:ex_val}
To demonstrate the effectiveness of the algorithm in a real-time application, we present an implementation on a test rig for a Vapor Compression Cycle (VCC). As explained in \cite{VCC_MB:2024}, heat transfer values for thermofluidic systems -- especially in the presence of phase changes -- tend to be very hard to determine and are usually dependent on process parameters. If the geometric dimensions of systems are known to a high degree, empirical relations exist in the literature, e.g. \cite{Liu:2018}, although these usually still come with large uncertainties. In  \cite{VCC_Part:2024}, we solved this by introducing integral feedback action through the heat-transfer value. To further increase the performance, we want to utilize the presented RGP-dKF algorithm to learn a model for the heat-transfer value and utilize this model afterwards for the nonlinear control published in \cite{VCC_Part:2024}. The implementation was carried out on a dedicated test rig at the Chair of Mechatronics at the University of Rostock. For details about the test rig please refer to \cite{VCC_MB:2024,VCC_Part:2024}.

\subsection{Physical Modeling}
The implementation of the RGP-dKF relies on a control-oriented system model. In this paper, we use a modified version of the lumped-parameter model presented in \cite{VCC_IOL:2023}.  By applying the simplification $\dot{m}_{C} \approx\dot{m}_{E} \approx \dot{m}_{m}=\frac{\dot{m}_{E}+\dot{m}_{C}}{2}$ , which is exact for equilibrium points due to conservation of mass, we reduce the two ODEs from \cite{VCC_IOL:2023} to a single ODE
\begin{equation}
\resizebox{1\hsize}{!}{$
\frac{dh_{Evap}}{dt}=\frac{\dot{m}_{m}  (h_{Evap,in}- h_{Evap,out})+K_{\alpha}\alpha \cdot A_{Evap} \cdot (\hat{T}_U-\hat{T}_{Evap}) }{V_{Evap} \rho_{Evap}}  \,.$
}
\end{equation}
Regarding the evaporator output enthalpy, we use a central discretization scheme
\begin{equation}
\label{eq:hout}
h_{Evap,out} = \frac{h_{Evap}-(1-\mathrm{w_\rho})h_{Evap,in}}{\mathrm{w_\rho}},  
\end{equation}
with  $\mathrm{w_\rho}=0.5$. In a nonlinear state-space representation with $x=h_{Evap}$, the simplified model becomes
\begin{equation}
\resizebox{1\hsize}{!}{$
\dot{x}=\frac{(h_{Evap,in}(t)- x)  \frac{\dot{m}_{m}(t)}{\mathrm{w_\rho}}+K_{\alpha}\alpha (t) \cdot A_{Evap} \cdot (\hat{T}_U (t)-\hat{T}_{Evap}(t)) }{V_{Evap} \rho_{Evap}(x,t)}\,,$
} 
\label{eq:Expred} 
\end{equation}
with the linear measurement output equation $y=h_{Evap,out}$, where we assume that $h_{Evap,out}=f(T_{Evap,out},p_{Evap,out})$ is directly measurable. Please note that, in contrast to  \cite{VCC_IOL:2023}, the original heat-transfer coefficient $\alpha$ is used as a normalization factor and $K_\alpha$ is identified instead. The values $\hat{T}_U,\dot{m}_{m},h_{Evap,in},\hat{T}_{Evap}$ represent measurable disturbances, which are assumed to be deterministic. Neglecting the enthalpy dependency of the density $\rho_{Evap}$, the time-varying partial derivative of the the right-hand side of \eqref{eq:Expred} w.r.t.~the state and the respective disturbance input gain become
\begin{equation}
a(t)=-\frac{\dot{m}_{m}(t)}{\mathrm{w_\rho} V_{Evap} \rho_{Evap}(t)} ,\, e(t)=\frac{\alpha  A_{Evap}  (\hat{T}_U (t)-\hat{T}_{Evap}(t)) }{V_{Evap} \rho_{Evap}(t)}\,,
\end{equation}
whereas the measured output factor is equal to the constant $h = \frac{1}{\mathrm{w_\rho}}$. 

To discretize the continuous-time model as needed for the EKF, we apply an explicit Euler discretization.

\subsection{RGP-dKF Implementation}
As previously explained, the goal of the RGP-dKF is to learn a GP model for the heat-transfer value of the evaporator. As explained in \cite{VCC_MB:2024}, there exist a number of dependencies that could qualify as meaningful inputs. The most important ones are the superheating enthalpy and the massflows of the refrigerant and coolant. Since the coolant massflow is not altered in the presented measurements, the envisaged GP model can be stated as $K_{\alpha,GP}= z_{GP}(u_C, \Delta h_{SH})$. Here, $u_C$ is used because it strongly correlates with the refrigerant mass flow. Moreover, it is -- in contrast to the massflow -- deterministic and an input of the partial IOL.
The implementation of the algorithm directly follows the description in Sec.~\ref{sec:KF_integration}. The trained GP model serves as an additive term w.r.t. the integrator state $K_{\alpha_I}$ of the $\alpha_I$-control in \cite{VCC_Part:2024} i.e. $K_{\alpha}= K_{\alpha,GP}+ K_{\alpha_I}$. During the evaluation of the trained GP model, the second input into the GP is replaced by the reference value, which aims to preserve the pure feedforward structure of this submodel w.r.t. the partial IOL from \cite{VCC_Part:2024}.

For the implementation, a Bachmann PLC (CPU: MH230) is used at the test rig with a sampling time of $T=10~\mathrm{ms}$. The chosen dimensions are $N_1=5$ for the first input $ \zeta_1=u_C$ and  $N_2=5$ for the second input  $ \zeta_2=\Delta h_{SH}$, which leads to a GP dimension of $N_1 N_2=25$ BVP. Please note that this value does not mark the computational limit because even GP dimensions up to $N_1 N_2=80$ BVP have been successfully tested.

\subsection{Experimental Results}
As the ground truth is not known, an experimental validation of the presented algorithm for the VCC tends to be a difficult task. Nevertheless, a usage of the learned GP model is related to a significantly improved performance, which suggests a general functionality.

\begin{figure}
	 \begin{center}
		 \includegraphics[width=1\linewidth]{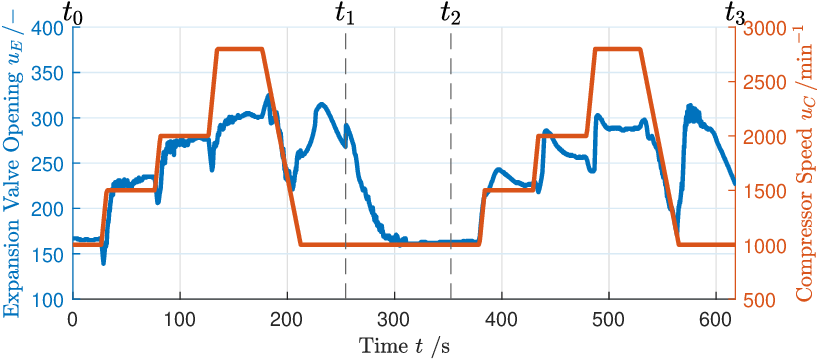}
		  \caption{Control inputs for the experimental validation of the RGP-dKF algorithm in VCC control.} 
		  \label{pic:PS_u}
	 \end{center}
\end{figure}

\begin{figure}
	 \begin{center}
		 \includegraphics[width=1\linewidth]{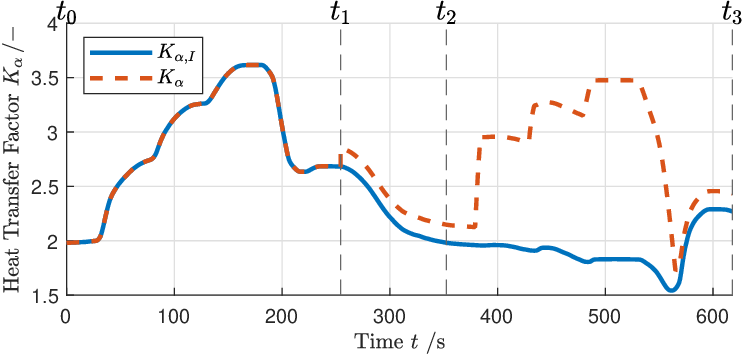}
		  \caption{$\alpha$-factor of the integrator feedback and combined $\alpha$-factor including the GP prediction for the experimental validation of the RGP-dKF algorithm in VCC control.} 
		  \label{pic:PS_K_alpha}
	 \end{center}
\end{figure}

As depicted in Fig. \ref{pic:PS_u}, we applied a step-wise rate-limited trajectory to the compressor speed of the system at time $t_0$. Moreover, the partial IOL control is employed to keep the tracking error of the superheating enthalpy $e_{\Delta h_{SH}}=\Delta h_{SH}-\Delta h_{SH,d}$  close to zero. As shown in Fig.~ \ref{pic:PS_K_alpha}, the integrator state, which covers the whole $K_\alpha$ during this phase ($t_0 \rightarrow t_1$), changes quite drastically to achieve this task. The learning of the RGP-dKF is enabled from $t_0$ to $t_1$. At time $t_1$, the learning is disabled and the GP prediction is added to the state of the integrator feedback. After a settling time  at time $t_2$, the same $u_C$-trajectory is restarted. Now, the learned part $K_{\alpha,GP}$ makes up a certain amount of $K_{\alpha}= K_{\alpha,GP}+ K_{\alpha_I}$, and the integrator state $K_{\alpha,I}$ changes much less in comparison with the first evaluation of the trajectory -- because most of the changes of $K_{\alpha}$ are now predicted by the learned GP model. Since this allows for a much quicker response of the $K_{\alpha}$, the control performance increases noticeably during the second evaluation of the trajectory. If the RMSE of the control output for the first evaluation of the $u_C$-trajectory  ($t_0 \rightarrow t_1$) is compared to the second one ($t_2 \rightarrow t_3$), a reduction of about $30\%$ can be stated along with a reduction of the necessary steps of the stepper motor of the expansion valve by more than $30\%$. This validates the application of the RGP-dKF algorithm for the given case. The general improvements turned out to be reproducible in different tests. Here, the quality of the GP prediction naturally improves with an increasing amount of data.

\section{Conclusions and Outlook}
This paper presents an innovative real-time capable algorithm for the learning of a Gaussian Process (GP) to cover unknown characteristics or relationships within a nonlinear system model. Here, also the estimation of not directly measurable states may be included. First, some modifications of an existing recursive Gaussian Process regression algorithm are proposed. Next, the seamless integration into an Extended Kalman Filter is presented for the combined state estimation and learning of a GP. The algorithm is validated and compared against two alternative implementations, including a concurrently published variant based on an auxiliary parameter. The proposed RGP-dKF algorithm was implemented on a test rig to learn a GP model for the heat-transfer value in the evaporator of a Vapor Compression Cycle. The GP model predictions were successfully used in a previously derived partial IOL control and could increase the control performance significantly, which corresponds to a successful experimental validation of the algorithm.

Since the alternative implementation based on the auxiliary parameter is yet unpublished, only selected performance measures are stated for a comparison. A thorough side by side comparison of the advantages and disadvantages of both variants is desirable. As many systems possess subsystems which are hard to describe by classical modeling approaches, the presented algorithm is interesting for many different applications. Further extensions of the algorithm might include the consideration of prior model knowledge about the hidden function including monotonicity, for which we already have some promising results. Finally, a learning of the remaining uncertainty of the GP prediction, which is so far modeled by a user-defined constant, could provide some further performance benefits.

\bibliography{ifacconf,myrefs2}

\end{document}